\documentclass[sigconf]{acmart}

\usepackage{subfigure}
\usepackage{multirow}
\usepackage{booktabs}
\usepackage{comment}
\AtBeginDocument{%
  }

\setcopyright{acmcopyright}
\copyrightyear{2023}
\acmYear{2023}
\acmDOI{XXXXXXX.XXXXXXX}

\begin{document}

\title{
Exploring Crowd Dynamics: Simulating Structured Behaviors through Crowd Simulation Models}


\author{Thiago Gomes Vidal de Mello}
\affiliation{%
  \institution{Virtual Humans Lab, PUCRS}
  \streetaddress{Av. Ipiranga, 6681}
  \city{Porto Alegre}
  \state{RS}
  \country{Brazil}
  \postcode{90619-900}
}
\email{thiago.mello.001@acad.pucrs.br}

\author{Matheus Schreiner Homrich da Silva}
\affiliation{%
  \institution{Virtual Humans Lab, PUCRS}
  \streetaddress{Av. Ipiranga, 6681}
  \city{Porto Alegre}
  \state{RS}
  \country{Brazil}
  \postcode{90619-900}
}
\email{matheus.silva00@edu.pucrs.br}

\author{Gabriel Fonseca Silva}
\affiliation{%
  \institution{Virtual Humans Lab, PUCRS}
  \streetaddress{Av. Ipiranga, 6681}
  \city{Porto Alegre}
  \state{RS}
  \country{Brazil}
  \postcode{90619-900}
}
\email{gabriel.fonseca94@edu.pucrs.br}

\author{Soraia Raupp Musse}
\affiliation{%
  \institution{Virtual Humans Lab, PUCRS}
  \streetaddress{Av. Ipiranga, 6681}
  \city{Porto Alegre}
  \state{RS}
  \country{Brazil}
  \postcode{90619-900}
}
\email{soraia.musse@pucrs.br}

\renewcommand{\shortauthors}{Redacted et al.}

\begin{abstract}
  This paper proposes the simulation of structured behaviors in a crowd of virtual agents by extending the BioCrowds simulation model. 
  Three behaviors were simulated and evaluated, a queue as a generic case and two specific behaviors observed at rock concerts. The extended model incorporates new parameters and modifications to replicate these behaviors accurately. Experiments were conducted to analyze the impact of parameters on simulation results, and computational performance was considered. 
  The results demonstrate the model's effectiveness in simulating structured behaviors and its potential for replicating complex social phenomena in diverse scenarios.
\end{abstract}




\begin{teaserfigure}
  \includegraphics[width=\textwidth]{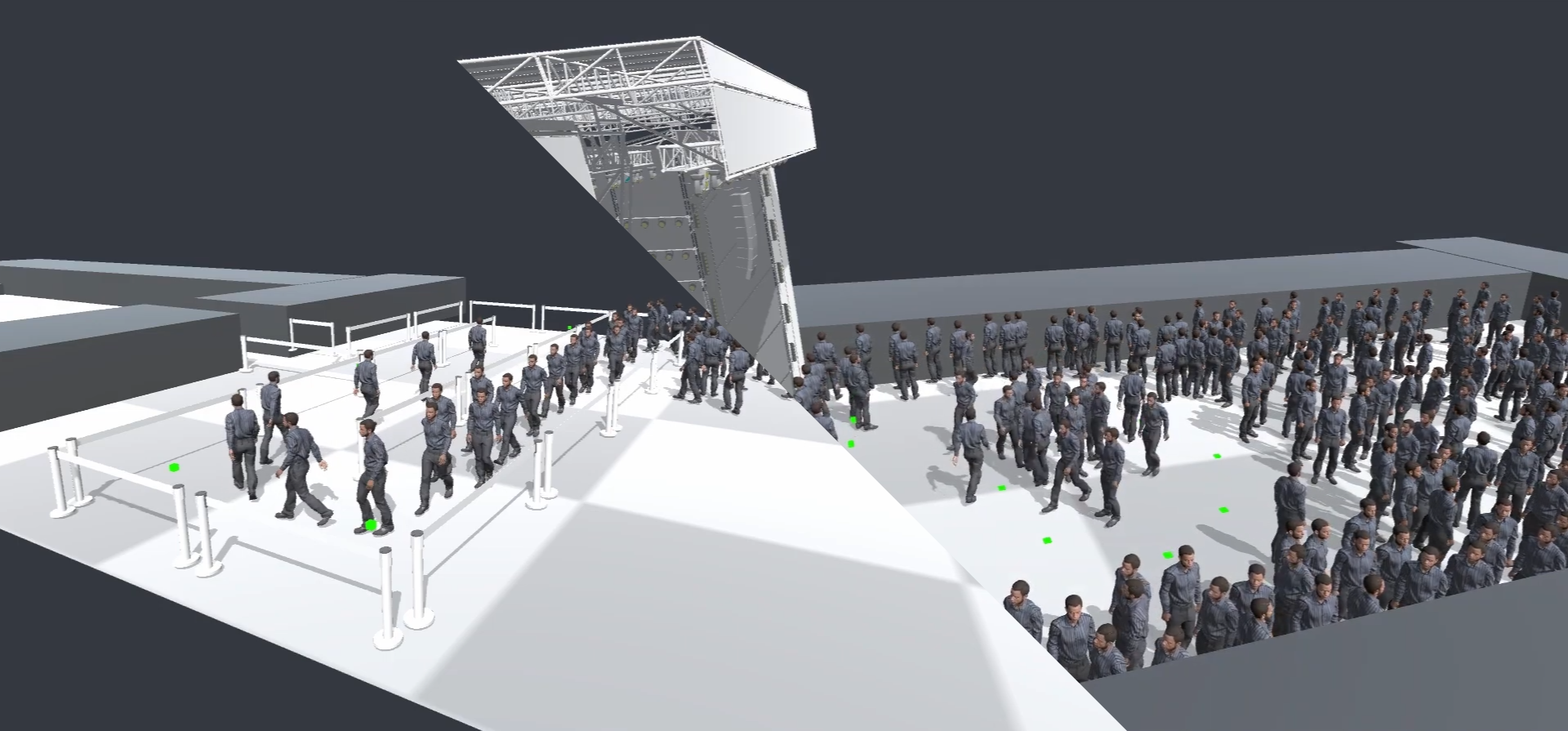}
  \caption{Simulation of structured behaviors during a rock concert. On the left, a group of agents enters the venue, forming an organized queue and moving in the available space between barriers. On the right, a group of agents participates in a \textit{moshpit} structured behavior while other agents create an open space for them.}
  \Description{Simulation of structured behaviors during a rock concert.}
  \label{fig:teaser}
\end{teaserfigure}


\newcommand\red[1]{{\color{red}#1}}

\maketitle

\section{Introduction}
\label{sec:introduction}

Since the introduction of the first crowd simulation model in computer graphics, proposed by Musse and Thalmann~\cite{10.1007/978-3-7091-6874-5_3}, significant advancements have been made to the field, enabling the simulation of a variety of scenarios and behaviors. This includes the simulation of diverse crowd behavior~\cite{sibgrapi_estendido}, crowd egress during evacuations~\cite{ReinforcedEvacuation}, personality traits in virtual agents\cite{durupinar2009ocean,knob2018simulating}, navigation control and path prediction\cite{Paris2007, silva2020fastForward}, and the comparison of crowds of high density~\cite{Pelechano:2007,Narain:2009}. 
In the entertainment industry, crowd simulation plays a significant role in controlling the behavior of Non-Playable Characters (NPCs), such as managing large armies and orchestrating their movement from one point to another while avoiding collisions with obstacles or other NPCs~\cite{suyikno2019feasible}.
Additionally, incorporating realistic and diverse group behaviors for virtual agents can significantly enhance the user's gaming experience, as indicated by studies on crowd perception. For instance, research has explored the perception of different character models within crowd simulations~\cite{mcdonnell2008clone}, the consideration of cultural features in virtual crowds~\cite{araujo2019much, araujo-cgi:2021}, perception of gender bias in virtual characters~\cite{araujo2022sigg}, and the examination of user interactions with virtual crowds in immersive virtual reality environments~\cite{volonte:2020}.

In this work, we aim to evaluate the capabilities of BioCrowds, a crowd simulation model proposed by Bicho et al~\cite{BICHO201270}, in simulating structured agent behavior in specific scenarios. 
BioCrowds was chosen as it offers mathematically proven collision-free simulations and has been successfully applied to simulate emergent behaviors in large populations~\cite{borges2017giving, knob2018simulating}.
Our proposal simulates three distinct structured behaviors in a crowd of virtual agents: a generic case representing organized and singular behavior (a queue) and two specific behaviors observed at rock concerts. 

It is essential to clarify the concept of structured behavior within the context of this work. While emergent behaviors arise spontaneously from the interactions between individuals, such as the formation of arches in front of doors during egress or the creation of lanes in high-density situations to optimize the flow of people~\cite{10.1007/978-3-7091-6874-5_3}, structured behaviors are predetermined by the population. These behaviors require understanding the population or a specific culture and how people typically behave, i.e., the population learns structured behaviors which are not emergent from people's interaction. Thus, while emergent behaviors occur naturally from individual interactions, the individuals learn and guide structured behaviors. The primary goal of this work is to determine the appropriate set of parameters for controlling a group of virtual agents using BioCrowds, enabling the simulation of desired structured behaviors in the given scenario. 

This paper is organized as follows: Section~\ref{sec:related_work} presents the related work regarding the simulation of agent behaviors in musical events and the analyses of real-world audience behaviors.
Section~\ref{sec:proposed_model} presents our proposed model for simulating structured behaviors in virtual agents using BioCrowds and a proposed set of parameters for each structured behavior.
Section~\ref{sec:results} presents the results achieved by our proposed model in simulated structured behaviors. 
Section~\ref{sec:final_considerations} presents our method's final considerations and future work.

\section{Related Work}
\label{sec:related_work}

The simulation of realistic musical events remains relatively limited in current research. One example is the work of Beacco et al.~\cite{direStraits}, which simulates a performance by the band Dire Straits based on recorded footage. However, this work primarily focuses on simulating the band itself, and the simulated audience does not accurately reflect the behavior of a real audience attending the performance. Another relevant work by Yılmaz et al.~\cite{musicDriven} aims to simulate a concert setting where the audience reacts to the music being played, resulting in more realistic behaviors throughout the performance. While the audience is not the central focus of this work, it contributes to the overall immersive experience. Additionally, a project focusing on public representation~\cite{vocaloid} has been developed to create realistic virtual reality simulations of shows featuring the virtual character Hatsune Miku from the Vocaloid software. This project replicates observed movements from real-world performances, providing valuable insights into constructing realistic simulation scenarios and identifying evaluation metrics.

Some methodologies have been proposed in previous work concerning simulation and observation of real crowds~\cite{Villamil2003}~\cite{Favaretto2016}~\cite{Favaretto2019}.
Regarding data relating to real-world audiences at concerts, an experiment was conducted at a music festival in Denmark, where Bluetooth signal sensors were employed to measure crowd density per square meter for different music genres presented at the event~\cite{bluetooth}. Although the parameterization of an audience has been explored, it has primarily focused on pop-rock concerts~\cite{poprock}, using live performances by artists to define parameters such as density, movement, and group dynamics. The data from these studies can serve as a valuable resource for parameterizing audience behavior in musical event simulations.

\subsection{BioCrowds}
\label{sec:biocrowds}

Regarding crowd simulation methods, the BioCrowds model~\cite{BICHO201270} utilizes markers distributed throughout the scenario to control the movement of agents. These markers have a capture area that determines their influence on an agent's movement direction. A collision-free behavior emerges from the presence of these markers, as they are not placed within obstacles such as walls, and each marker is associated with a single agent, specifically the one closest to it. Additionally, BioCrowds has been applied with various parameterized behaviors beyond its original design. For instance, Rockenbach~\cite{10.1145/3267851.3267872} proposed a model incorporating comfort and panic parameters, influencing agent density in specific simulation areas. Another study focused on agents navigating a flooded area, where water influenced their movement~\cite{flood}.

To generate markers in the scenario, BioCrowds employs the Dart Throwing algorithm~\cite{DartThrowing}, which randomly distributes markers. These markers serve as reference points for agents during the simulation. Each agent possesses a target point in the scene towards which it aims to navigate (i.e., a goal). Furthermore, agents have a capture area defined by a radius. Within this radius, agents capture markers, and the positions of these captured markers determine the agent's movement direction. The calculation of the movement direction considers the weights assigned to the markers, with markers in the goal's direction carrying higher weights than others, ranging from 1 to 0.

Table~\ref{tab:parameters} displays the parameters available in the original BioCrowds model. The table includes the following parameters: $MaxAgents$, which represents the maximum number of agents in the scene; $AgentRadius$, which denotes the marker capture radius for each agent; $MarkerDensity$ and $MarkerRadius$, representing the density and radius of markers in the environment, respectively, and are used by the Dart Throwing algorithm~\cite{DartThrowing}; $GoalList$, which specifies the list of goals that each agent will follow; and $GoalDistanceThreshold$, indicating the proximity threshold at which agents consider they have reached their goal and proceed to the next one in the list.
There are also parameters to control the creation of new agents during the simulation. Each spawn area contains: $SpawnAreaInitialAgents$, which represents the number of agents created at the start of the simulation; $SpawnAreaGoalList$ represents the goals list for agents created by this spawn area; and $QuantitySpawnedEachCycle$ which denotes the number of agents instantiated after each cycle, defined by $CycleLength$ seconds.
Additionally, the table presents the proposed parameters for simulating structured behaviors, which are detailed in Section~\ref{sec:proposed_model}.

\begin{figure}[!ht]
\centering
\includegraphics[width=0.48\textwidth]{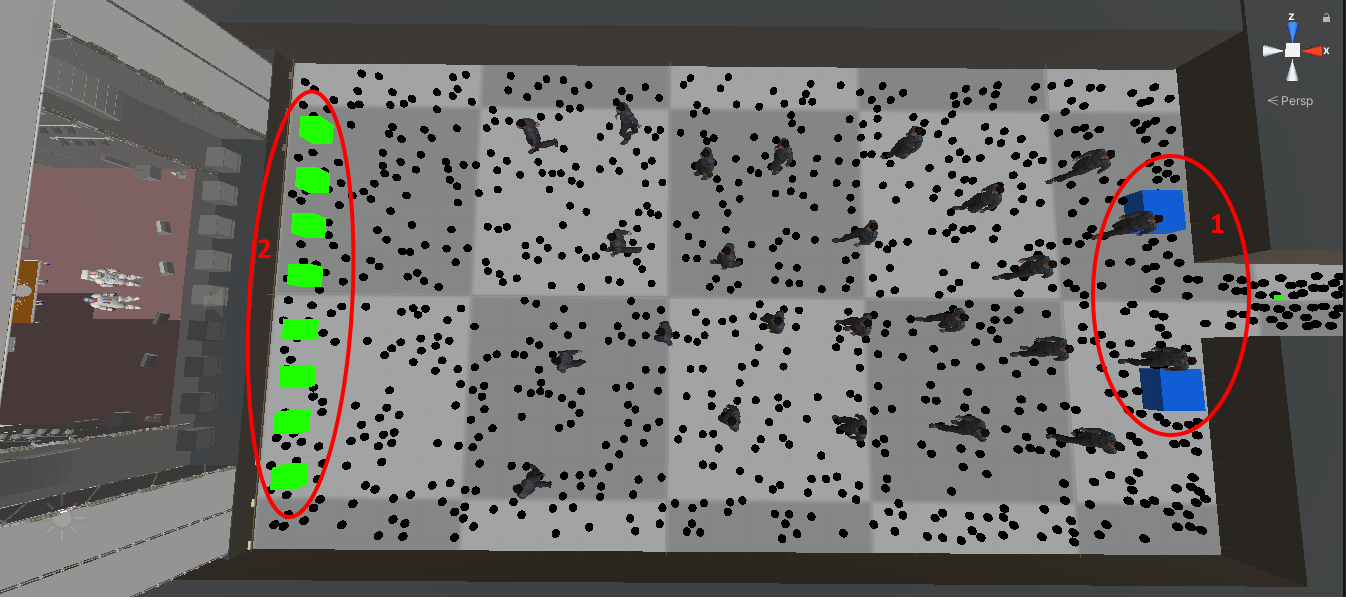}
\caption{Image representing an initial step of a simulation, with virtual agents moving from right to left. The spawn areas of agents are represented by blue cubes, highlighted within ellipse 1. The possible goals for agents are represented by green cubes, highlighted within ellipse 2. Markers, utilized in the model to facilitate agent movement, are represented as black dots.}
\label{fig:biocrowds_example}
\end{figure}

Figure~\ref{fig:biocrowds_example} illustrates an initial step during a simulation, with black dots representing markers generated by the Dart Throwing algorithm~\cite{DartThrowing}. Agents can be observed moving from their spawn areas (i.e., the location where agents are created) on the right side of the image, denoted by the blue cubes highlighted within ellipse 1, towards their designated goal on the left side, indicated by the green cubes within ellipse 2.

\section{Proposed Model}
\label{sec:proposed_model}

In this study, we aim to evaluate the capabilities of BioCrowds~\cite{BICHO201270} in simulating structured behaviors in virtual agents, focusing on testing parameters to represent individuals and their behavior in a specific scenario accurately. Specifically, we aim to simulate the audience of a musical event featuring metal bands, incorporating structured behaviors commonly observed in such events. 
The selected behaviors for simulation are \textit{moshpit}, a \textit{circlepit}, and a queue. The first two behaviors are distinct and typically observed at rock concert events, while the queue behavior represents a more generic scenario that may occur in different locations. Detailed descriptions of these three structured behaviors will be provided in the following subsections. 
Table~\ref{tab:parameters} present the parameters required for simulating the proposed structured behaviors alongside a short description. The table also presents the parameters already included in the original BioCrowds models~\cite{BICHO201270}, as described in Section~\ref{sec:biocrowds}.

\begin{table*}[htp]
\centering
\begin{tabular}{@{}clcccc@{}}
\toprule
Parameter                   & \multicolumn{1}{c}{Description}                                                                                                                                      & \begin{tabular}[c]{@{}c@{}}Original\\ BioCrowds\end{tabular} & Moshpit & Circlepit & Queue \\ \midrule
Max Agents                  & Maximum number of agents in the scene.                                                                                                                               & \checkmark                                                        & \checkmark   & \checkmark     & \checkmark \\ \midrule
Agent Radius                & Agent radius for capturing markers.                                                                                                                                  & \checkmark                                                        & \checkmark   & \checkmark     & \checkmark \\ \midrule
Marker Density              & \begin{tabular}[c]{@{}l@{}}Density of markers in the simulation scene, used by\\ the Dart Throwing algorithm~\cite{DartThrowing}.\end{tabular}                                                                           & \checkmark                                                        & \checkmark   & \checkmark     & \checkmark \\ \midrule
Marker Radius               & \begin{tabular}[c]{@{}l@{}}Radius of markers in the simulation scene, used by\\ the Dart Throwing algorithm~\cite{DartThrowing}.\end{tabular}   & \checkmark                                                        & \checkmark   & \checkmark     & \checkmark \\ \midrule
Goal List                   & \begin{tabular}[c]{@{}l@{}}List of goals that each individual agent will follow\\during the simulation.\end{tabular}  & \checkmark                                                        & \checkmark   & \checkmark     & \checkmark \\ \midrule
Goal Distance Threshold     & \begin{tabular}[c]{@{}l@{}}Proximity threshold for agents to consider reaching \\ their current goal.\end{tabular}                                                      & \checkmark                                                        & \checkmark   & \checkmark     & \checkmark \\ \midrule
Goal Wait List & \begin{tabular}[c]{@{}l@{}}Agent wait time between reaching a goal and targeting\\ the next one.\end{tabular}  & \checkmark                                                        & \checkmark   & \checkmark     & \checkmark \\ \midrule
Spawn Area Initial Agents    & \begin{tabular}[c]{@{}l@{}}Number of agents instantiated by a spawn area at\\ the beginning of the simulation.\end{tabular}                                             & \checkmark                                                        & \checkmark   & \checkmark     & \checkmark \\ \midrule
Spawn Area Goal List & \begin{tabular}[c]{@{}l@{}}Reference to goals in the scene, serving as targets \\ for agents created by a spawn area.\end{tabular} & \checkmark                                                        & \checkmark   & \checkmark     & \checkmark \\ \midrule
Cycle Length & \begin{tabular}[c]{@{}l@{}}Delay in seconds for a spawn area to instantiate\\ new agents in the simulation.\end{tabular}                                                       & \checkmark                                                        & \checkmark   & \checkmark     & \checkmark \\ \midrule
Quantity Spawned Each Cycle & Number of agents generated, per cycle, in a spawn area.                                                                                                                                & \checkmark                                                        & \checkmark   & \checkmark     & \checkmark \\ \midrule

Number Agents Pit           & \begin{tabular}[c]{@{}l@{}}Number of agents participating in the \textit{moshpit} and\\ \textit{circlepit} behaviors.\end{tabular}                                                     &                                                              & \checkmark   & \checkmark     &       \\ \midrule
Area of Effect Radius       & \begin{tabular}[c]{@{}l@{}}Radius of the area of effect used in the \textit{moshpit} and\\\textit{circlepit} behaviors, represented as a sphere.\end{tabular}  &                                                              & \checkmark   & \checkmark     &       \\ \midrule
Moshpit Center Goal           & \begin{tabular}[c]{@{}l@{}}Reference to the center goal of the area of effect, added\\ to the goal list of agents participating in the behavior.\end{tabular}        &                                                              & \checkmark   &    \checkmark       &       \\ \midrule
Circlepit Goal List         & \begin{tabular}[c]{@{}l@{}}Reference to the goals inside the area of effect, added  \\ to the goal list for agents participating in the \textit{circlepit}.\end{tabular}                &                                                             &         & \checkmark     &       \\ \midrule
Reflect Threshold (Min/Max) & Distance limits for the agent repulsion behavior.      &                                                           & \checkmark   &     &       \\ \midrule
Time To Start            & \begin{tabular}[c]{@{}l@{}}Delay in seconds to start moving agents to the \\ center of the \textit{moshpit} after creating the space.\end{tabular}                            &                                                              & \checkmark   & \checkmark     &       \\ \bottomrule
\end{tabular}
\caption{Description of the parameters used in the simulations. The table presents the parameters available in the original BioCrowds model~\cite{BICHO201270}, alongside the parameters required for simulating each proposed structured behavior.}
\label{tab:parameters}
\end{table*}

\subsection{Moshpit}
\label{sec:moshpit}

In this work, we define a \textit{moshpit} as a behavior that occurs within a designated space in the middle of the audience when many individuals move away from a central point. This space serves as a stage for various audience actions, including the \textit{moshing}, which involves individuals near the \textit{moshpit} launching themselves towards its center in an attempt to push others.

\begin{figure}[ht]
\centering
\includegraphics[width=0.45\textwidth]{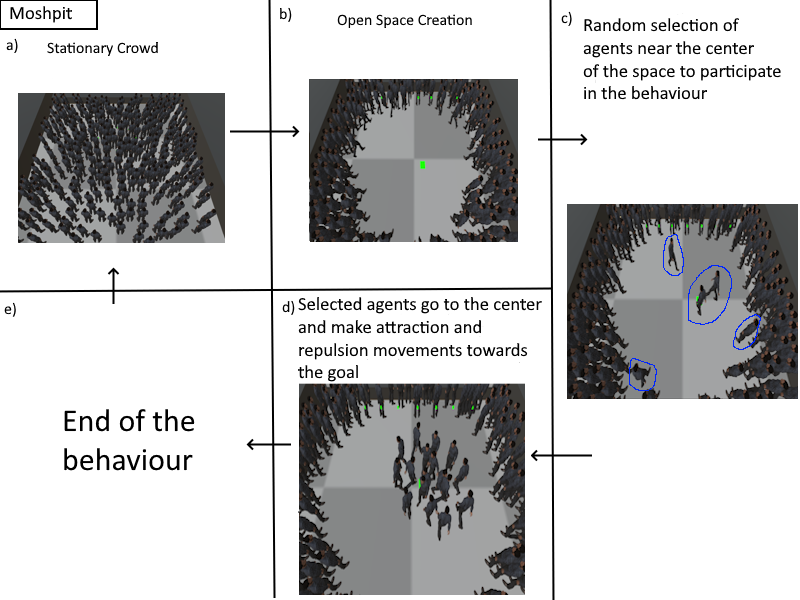}
\caption{The simulation steps for the \textit{moshpit} structured behavior. In (a), the initial state of the crowd is presented, before the behavior begins. In (b), the crowd creates an open space to initiate the behavior. In (c), agents near to the central area are randomly selected to participate in the behavior. In (d), the \textit{moshpit} behavior is executed, with agents moving and interacting accordingly in an alternating pattern of attraction and repulsion. Finally, in (e), the \textit{moshpit} behavior is concluded and the crowd returns to its initial state.}
\label{fig:moshpit_steps}
\end{figure}

Figure~\ref{fig:moshpit_steps} illustrates the planned steps for simulating a \textit{moshpit} structured behavior with virtual agents. In Step A, the agents are initially stationary in the scenario, awaiting the initiation of the behavior while observing the stage. In Step B, a designated space is created among the agents, determined by parameterized distances. From this space, a random selection of agents closest to the center is made in Step C. Subsequently, in Step D, these selected agents move towards the center, engaging in an alternating pattern of attraction and repulsion with respect to the central area. This process is repeated until the behavior is interrupted in Step E, at which point the agents return to their initial formation as part of the audience at the show.

\begin{figure}[ht]
\centering
\includegraphics[width=0.45\textwidth]{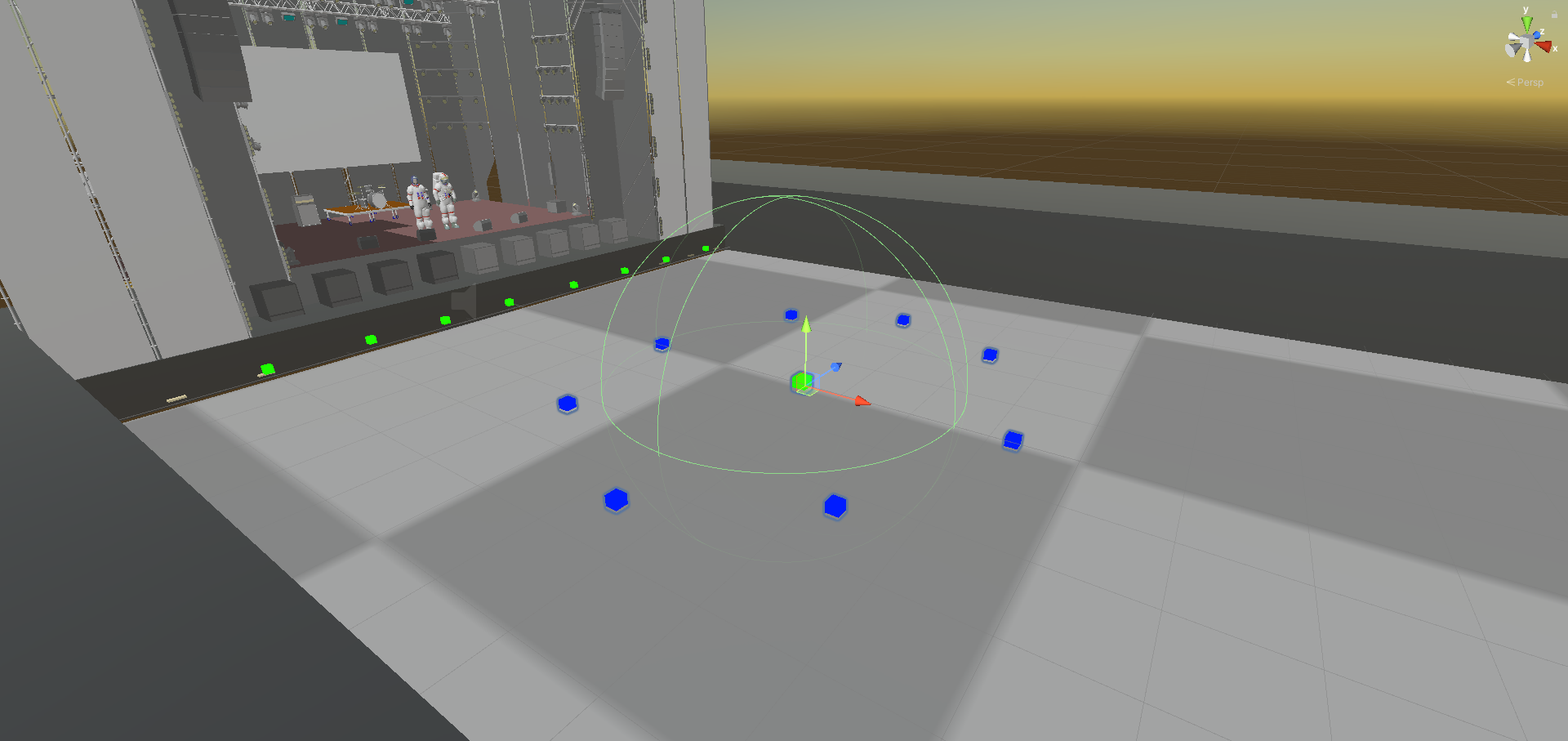}
\caption{Simulation scene containing an area of effect for the \textit{moshpit} and \textit{circlepit} structured behaviors, represented as a green wireframe sphere. The center of the sphere contains a goal, represented by a green cube, to repel agents when creating an open space at the beginning of the behavior. Additionally, eight goals, represented by blue circles, are placed near the area of effect and are used for the circular motion during the \textit{circlepit}.}
\label{fig:area_of_affect}
\end{figure}

In order to simulate the \textit{moshpit} behavior, several additions were made to the BioCrowds~\cite{BICHO201270} model. One of the key additions is the inclusion of an area of effect used to facilitate the execution of the\textit{moshpit} behavior. In this work, we define this area as a sphere.
Figure~\ref{fig:area_of_affect} illustrates the area of effect as a green sphere, indicating the designated location within the simulation scene where the \textit{moshpit} will take place. The center of the area contains a single goal, used to repel agents when creating an open space at the beginning of the behavior.

In our current implementation, the \textit{moshpit} behavior is initiated through user input. However, other options could be implemented.
When the behavior is initiated, the simulation will follow the steps depicted in Figure~\ref{fig:moshpit_steps}.
Using the area of effect, represented as a sphere, agents within the radius of this sphere are randomly chosen to participate in the \textit{moshpit}. These selected agents are assigned to the center of the sphere as their new goal, leading to a modification in the calculation of their marker weights. Specifically, instead of using the original weight value, we use 1 minus the original weight.
This alteration creates a repulsion behavior from a specific position, which was not present in the original BioCrowds model~\cite{BICHO201270}. It instructs the agents to move in the opposite direction of their goal, instead of towards it, resulting in the creation of empty space around the target at the center of the sphere. This repulsion behavior vacates the nearby space, making the markers available for later stages of the behavior.

Following the repulsion phase, a subset of agents closest to the center of the sphere is randomly selected. The number of agents participating in these selections is determined by the user prior to starting the simulation via the $NumberAgentsPit$ parameter. The chosen agents are then directed toward their new goal at the center of the sphere.
Subsequently, the agents alternate between moving towards the goal (center of the sphere) and being repelled from it.
This alternating movement attempts to replicate the mosh experience, where participating individuals push against each other. However, in this simulation, only the positional changes of the agents are considered, without actual interaction between them to generate the pushing effect. Table~\ref{tab:parameters} presents the parameters required for simulating the \textit{moshpit} structured behavior. 

\subsection{Circlepit}
\label{sec:circlepit}

We define \textit{circlepit} as a structured behavior that occurs within the open space in the middle of the audience, similar to the previous behavior. During a \textit{circlepit}, individuals within the space start to run in a circular line, creating a circle-like motion.

\begin{figure}[ht]
\centering
\includegraphics[width=0.45\textwidth]{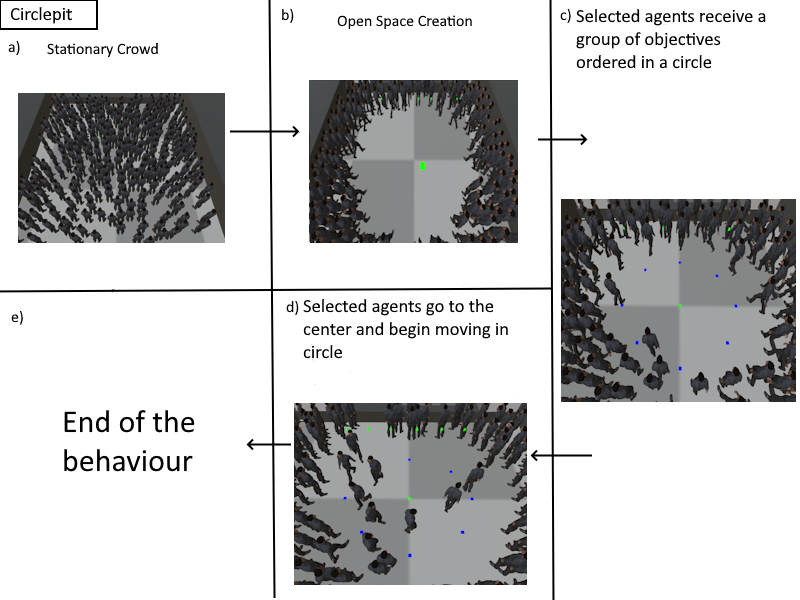}
\caption{The simulation steps for the \textit{circlepit} structured behavior. In (a), the initial state of the crowd is depicted before the behavior begins. In (b), the crowd opens up space to initiate the behavior. In (c), the goals that agents will use to carry out the behavior are highlighted in green cubes. In (d), the behavior is illustrated as agents move in a circular pattern within the designated area. Finally, in (e), the \textit{circle} behavior concludes and the crowd returns to its initial state.}
\label{fig:circlepit_steps}
\end{figure}

Figure~\ref{fig:circlepit_steps} illustrates the planned steps for executing the \textit{circlepit} behavior with the simulation agents. Step A shows the agents stationary in the scenario, awaiting the initiation of the behavior. Step B involves opening up a designated area within the crowd, creating a space for the behavior to take place. In step C, a random selection is made from the agents closest to the center of the open area. These chosen agents then engage in a circular movement within the designated space in step D. The circular motion continues until the behavior is interrupted in step E, at which point the agents return to their initial positions.

For this behavior, the steps up to step B of Figure~\ref{fig:circlepit_steps} remain the same as the previous behavior. However, a change is introduced in step C. To simulate the \textit{circlepit} behavior, agents need to follow a sequential list of goals that will be repeated while the behavior is active.
At the start of the behavior, participating agents receive a list of goals to follow, exclusive to the \textit{circlepit} behavior. These goals are represented by the blue cubes in Figure~\ref{fig:area_of_affect}. Then, agents are directed to the next goal in the list, considering their proximity. This allows for the repetition of goals from the agent's list, which has already been implemented in the BioCrowds model.
Table~\ref{tab:parameters} presents the parameters required for simulating the \textit{ciclepit} structured behavior. 

\subsection{Queue}
\label{sec:queue}

\begin{figure}[!htb]
  \centering
  \subfigure[fig:queuescene1][Queue Scene 1. No obstacles are present and agents must navigate through a sequence of goals.]{\includegraphics[width=0.48\textwidth]{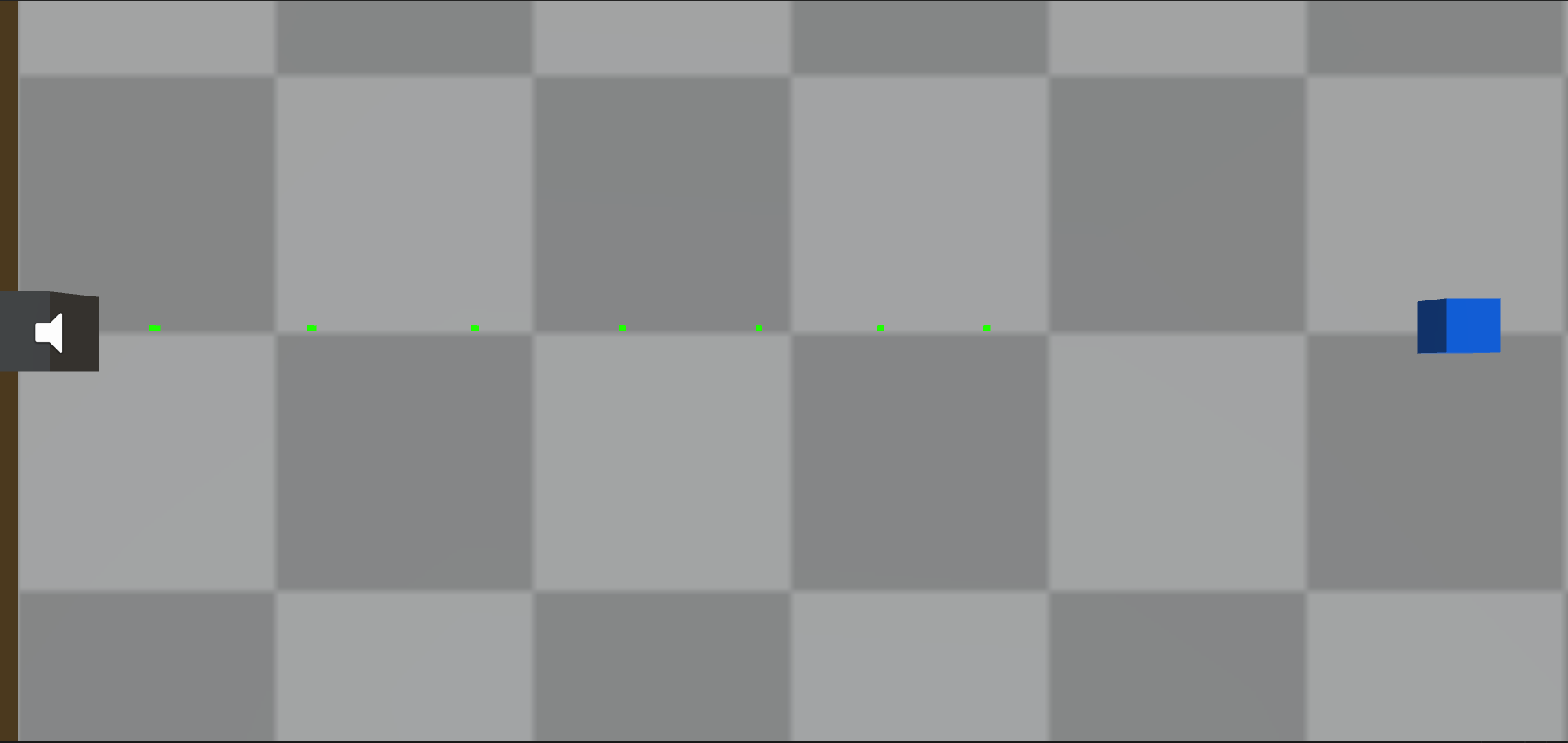}
  \label{fig:queue_scene_1}}
  \subfigure[fig:queuescene2][Queue Scene 2. Obstacles are present and agents have a single goal. Obstacles have a wide distance on the left and a narrow distance on the right.]{\includegraphics[width=0.48\textwidth]{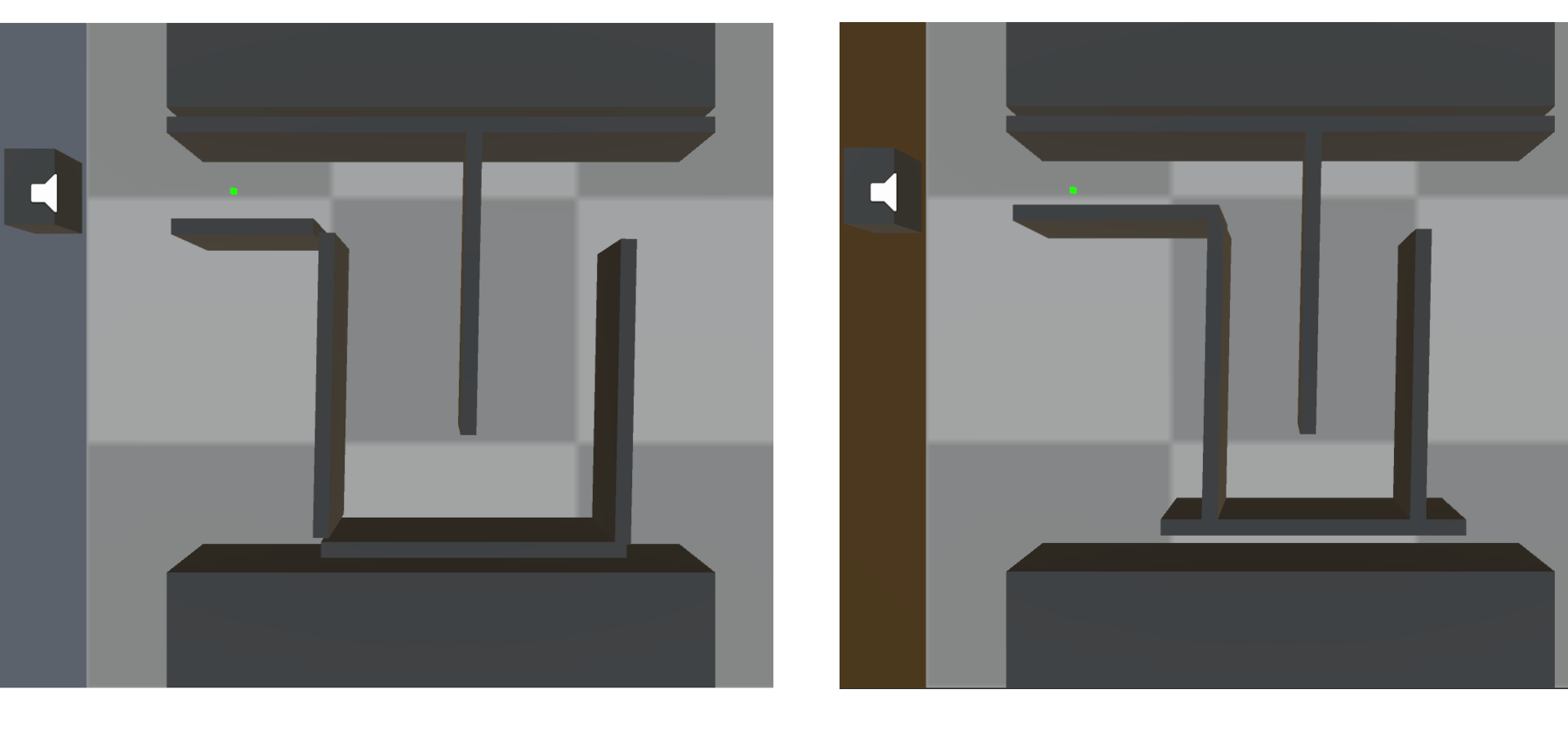}
  \label{fig:queue_scene_2}}
  \subfigure[fig:queuescene3][Queue Scene 3. Obstacles are present and agents must navigate through a sequence of goals. Obstacles have a wide distance on the left and a narrow distance on the right.]{\includegraphics[width=0.48\textwidth]{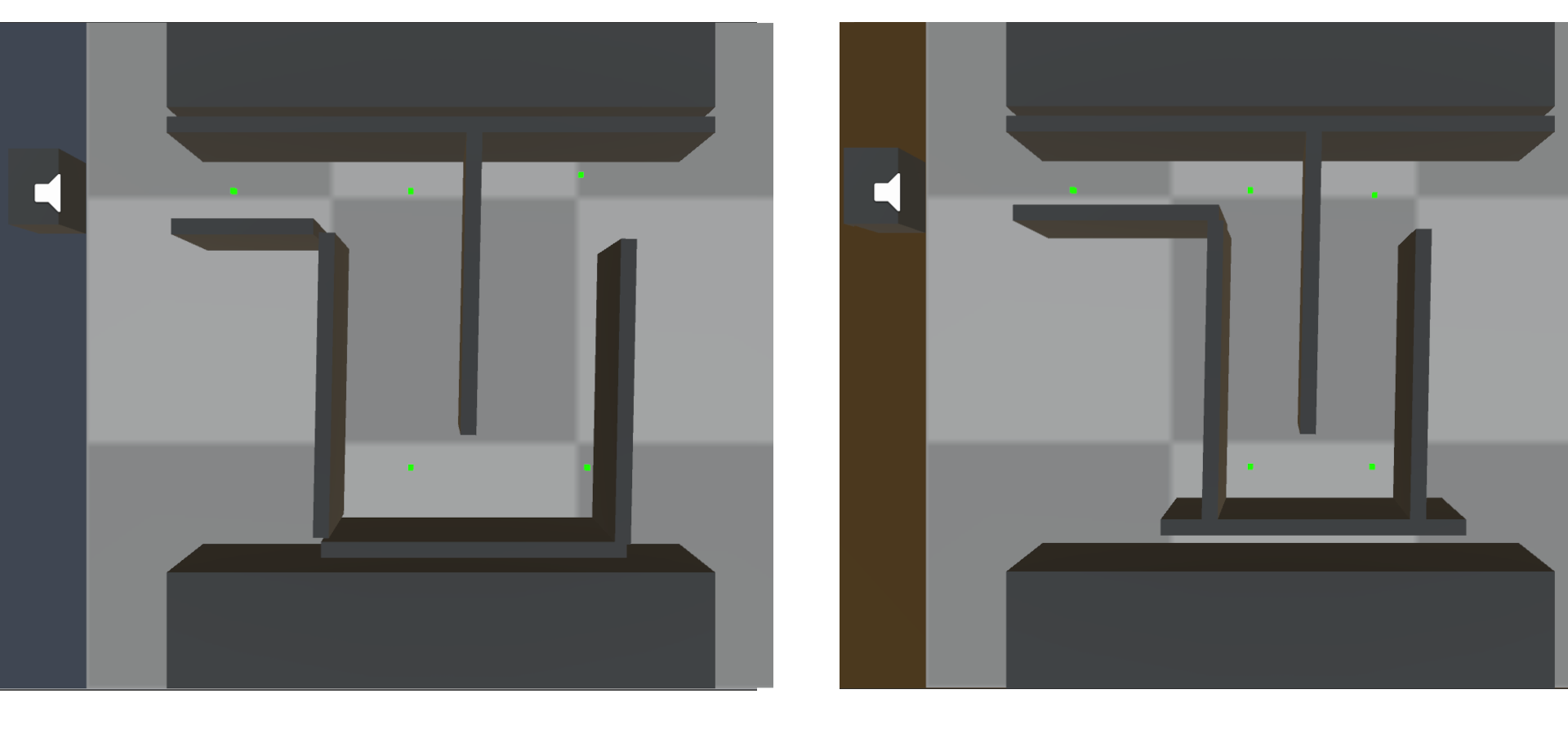}
  \label{fig:queue_scene_3}}
    \caption{Simulation scenes built for the queue experiments. In all scenarios, agents move from right to left. Agent goals are represented by green cubes within the available space. An agent must be within a distance threshold of their current goal before starting to move toward their next goal. }
    \label{fig:queue_scenes}
\end{figure}

The queue behavior is a nice example of learned behavior present in the social convention. The behavior is described by a number of people who compete for a certain resource. In order to simulate this behavior, we employed two methods. 
The first method aimed to assess the behavior of the original BioCrowds model and determine its suitability for simulating queues. The second method involved restricting the available space for agents using obstacles and guiding them through the queue using a sequence of goals. No modifications were required in the original model for this particular behavior, as presented in Table~\ref{tab:parameters}.
We conducted experiments in three distinct scenarios, presented in Figure~\ref{fig:queue_scenes}. The first scenario features a sequence of goals, where the agents can move towards the next goal once the current one is reached. In Figure~\ref{fig:queue_sim_scene1} it is possible to see how BioCrowds simulates the situation of a queue, i.e., it is not organized, as we cognitively expect, and the visual behavior does not present the clear sequence of agents to be served. In other words, our hypothesis is that the behavior illustrated in Figure~\ref{fig:queue_sim_scene1} should be the emergent and natural behavior of people when competing for a resource. The queue, as we expect, is a learned behavior.

The second scenario consists of obstacles along with a single goal placed at the end. The third scenario incorporates both obstacles and a sequence of goals placed within the available space. For the second and third scenarios, the experiments were conducted both with a wide and a narrow distance between obstacles, where a narrow space means that the area between obstacles is reduced and there is less available space for agents. The goal of such space restriction is to evaluate if agents will evolve the queue behavior without any new heuristics in the agents' intelligence, as made for the other two behaviors.

\section{Experimental Results}
\label{sec:results}

In this section, we present the results of the experiments conducted on the three proposed structure behaviors, as discussed in lasts sections. For each behavior, we provide a table showcasing the tested parameters, their values, and images generated from the respective simulations.

Regarding the \textit{moshpit} and \textit{circlepit} structured behaviors, we conducted experiments by varying the number of markers in the scenario and the marker capture range of agents, i.e., the $MarkerDensity$ and $AgentRadius$ parameters presented in Table~\ref{tab:parameters}.

Table~\ref{tab:mosh_circle_params} presents the parameter values used for the \textit{moshpit} and \textit{circlepit} simulations, along with the corresponding figures. We used $MarkerDensity$ values of $0.75$ and $0.5$, as well as $AgentRadius$ values of $1$ and $5$.
The maximum number of agents (i.e., $MaxAgents$) was defined as 300 for simulations with a $MarkerDensity = 0.75$, and 200 for simulations with a $MarkerDensity = 0.5$. In the latter case, there were fewer markers in the environment, resulting in fewer agents occupying the available space.
Finally, for all experiments, we defined $NumberAgentsPit = 20$, meaning that the desired number of agents to participate in the behavior is 20. However, the actual number of agents involved in the movement could vary due to random selection among nearby agents and potential obstruction caused by agents who did not participate. 

\begin{figure}[!htb]
  \centering
  \subfigure[fig:mosh1][Simulation of the \textit{moshpit} behavior considering $MaxAgents = 300$, $MarkerDensity = 0.75$, $AgentRadius = 1$.]{\includegraphics[width=0.42\textwidth]{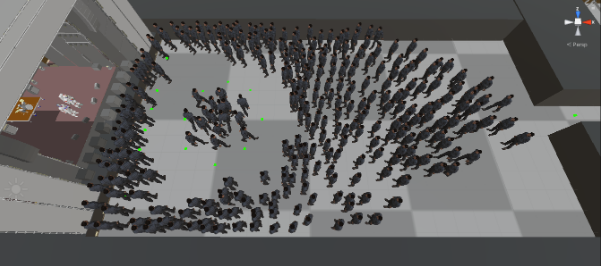}
  \label{fig:moshpit_1}}
  \subfigure[fig:mosh4][Simulation of the \textit{moshpit} behavior considering $MaxAgents = 200$, $MarkerDensity = 0.5$, $AgentRadius = 5$.]{\includegraphics[width=0.42\textwidth]{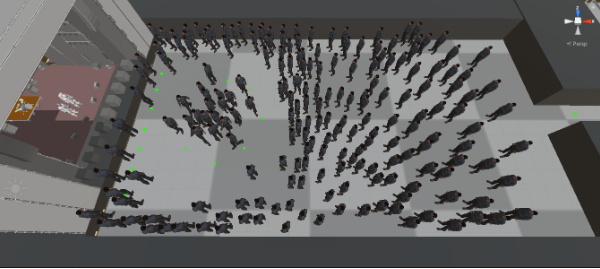}
  \label{fig:moshpit_4}}
    \caption{Simulations of \textit{moshpit} behavior, considering the parameter configurations presented in Table~\ref{tab:mosh_circle_params}.}
    \label{fig:moshpit_simulation}
\end{figure}

\begin{figure}[!htb]
  \centering
  \subfigure[fig:circle1][Simulation of the \textit{circlepit} behavior considering $MaxAgents = 300$, $MarkerDensity = 0.75$, $AgentRadius = 1$.]{\includegraphics[width=0.42\textwidth]{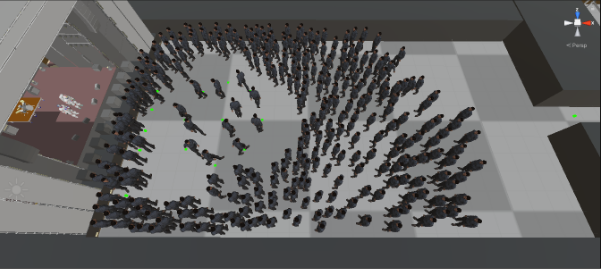}
  \label{fig:circlepit_1}}
  \subfigure[fig:circle4][Simulation of the \textit{circlepit} behavior considering $MaxAgents = 200$, $MarkerDensity = 0.5$, $AgentRadius = 5$.]{\includegraphics[width=0.42\textwidth]{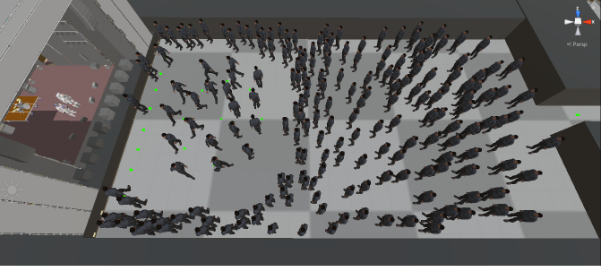}
  \label{fig:circlepit_4}}
    \caption{Simulations of \textit{circleplit} behavior, considering the parameter configurations presented in Table~\ref{tab:mosh_circle_params}.}
    \label{fig:circle_simulation}
\end{figure}

In the \textit{moshpit} simulations, the number of agents participating in the behavior varied between 16 agents (Figure~\ref{fig:moshpit_1}) and 20 agents (Figure~\ref{fig:moshpit_4}). 
In the \textit{circlepit} tests, we observed a greater variation, ranging from 12 agents (Figure~\ref{fig:circlepit_1}) to 19 agents (Figure~\ref{fig:circlepit_4}). 
We found that using the parameters $MarkerDensity = 0.5$ and $AgentRadius = 5$ resulted in more agents participation for both behaviors since the simulation will have more free space, because of the lower quantity of markers, as well as the agents being able to use farther markers due to the higher value of the parameter $AgentRadius$.

Video recordings of the \textit{moshpit} and \textit{circlepit} simulations are available online~\footnote{A playlist of the \textit{moshpit} simulations is available at: \\~\url{https://www.youtube.com/playlist?list=PLsk4Dh1ALO5nzIYk_6cr0esln7hbNR3fF} \\
A playlist of the \textit{circlepit} simulations is available at: \\~\url{https://www.youtube.com/playlist?list=PLsk4Dh1ALO5kZ0feRS-RDrnQo7f3PsE6k}
}.

\begin{table}[ht]
 \centering
\begin{tabular}{@{}lcccc@{}}
\toprule
                           & \begin{tabular}[c]{@{}c@{}}Max\\ Agents\end{tabular} & \begin{tabular}[c]{@{}c@{}}Marker\\ Density\end{tabular} & \begin{tabular}[c]{@{}c@{}}Agent\\ Radius\end{tabular} & Figure                                                                        \\ \midrule
\multirow{4}{*}{Moshpit}   & 300                                                  & 0.75                                                     & 1                                                      & Figure~\ref{fig:moshpit_1}                       \\ \cmidrule(l){2-5} 
                           & 300                                                  & 0.75                                                     & 5                                                      & ---                                                                           \\ \cmidrule(l){2-5} 
                           & 200                                                  & 0.5                                                      & 1                                                      & ---                                                                           \\ \cmidrule(l){2-5} 
                           & \textbf{200}                                                  & \textbf{0.5}                            & \textbf{5}                            & \textbf{Figure~\ref{fig:moshpit_4}}   \\ \midrule
\multirow{4}{*}{Circlepit} & 300                                                  & 0.75                                                     & 1                                                      & Figure~\ref{fig:circlepit_1}                       \\ \cmidrule(l){2-5} 
                           & 300                                                  & 0.75                                                     & 5                                                      & ---                                                                           \\ \cmidrule(l){2-5} 
                           & 200                                                  & 0.5                                                      & 1                                                      & ---                                                                           \\ \cmidrule(l){2-5} 
                           & \textbf{200}                                                  & \textbf{0.5}                            & \textbf{5}                         & \textbf{Figure~\ref{fig:circlepit_4}} \\ \bottomrule
\end{tabular}
\caption{Comparison of parameters used in the simulations of \textit{moshpit} and \textit{circlepit} behaviors. Values in bold represent the values deemed most suitable for simulating both structured behaviors.}
\label{tab:mosh_circle_params}
\end{table}

Next, we will discuss the results obtained with the simulation of the queue structured behavior. Table~\ref{tab:queue_params} presents parameter values and corresponding simulation images. The three distinct queue simulation scenes were described in Section~\ref{sec:queue} and are presented in Figure~\ref{fig:queue_scenes}. Similar to the \textit{moshpit} and \textit{circlepit} experiments, we used $MarkerDensity$ values of $0.75$ and $0.5$, as well as $AgentRadius$ values of $1$ and $5$. The Obstacle Distance column indicates if an experiment has a wide or narrow space between obstacles. By manipulating these key parameters, we aimed to enhance the realism of the simulation. Adjusting the density of agents in the space, modifying the movement range by expanding or restricting it, and introducing restricted areas allowed us to observe significant changes in the simulation outcomes.

Video recordings of the queue behavior simulations are available online~\footnote{
A playlist of the queue simulations is available at: ~\url{https://www.youtube.com/playlist?list=PLsk4Dh1ALO5kmnbSv4A6c5e1FYwSLO5me}
}.

\begin{table}[ht]
\centering
\begin{tabular}{@{}cccccc@{}}
\toprule
                                                                         & \begin{tabular}[c]{@{}c@{}}Number\\ of Goals\end{tabular} & \begin{tabular}[c]{@{}c@{}}Marker\\ Density\end{tabular} & \begin{tabular}[c]{@{}c@{}}Agent\\ Radius\end{tabular} & \begin{tabular}[c]{@{}c@{}}Obstacle\\ Distance\end{tabular} & Figures \\ \midrule
\begin{tabular}[c]{@{}c@{}}Queue\\ Scene 1\end{tabular}                  & 7                                                         & 0.75                                                     & 1                                                      & ---                                                         & Figure~\ref{fig:queue_sim_scene1}  \\ \midrule
\multirow{4}{*}{\begin{tabular}[c]{@{}c@{}}Queue\\ Scene 2\end{tabular}} & \multirow{4}{*}{1}                                        & 0.75                                                     & 1                                                      & Wide                                                         & Figure~\ref{fig:queue_sim_scene2_1}  \\ \cmidrule(l){3-6} 
                                                                         &                                                           & 0.75                                                     & 1                                                      & Narrow                                                        & Figure~\ref{fig:queue_sim_scene2_2}  \\ \cmidrule(l){3-6} 
                                                                         &                                                           & 0.5                                                      & 5                                                      & Wide                                                         & Figure~\ref{fig:queue_sim_scene2_3}  \\ \cmidrule(l){3-6} 
                                                                         &                                                           & 0.5                                                      & 5                                                      & Narrow                                                        & Figure~\ref{fig:queue_sim_scene2_4}  \\ \midrule
\multirow{4}{*}{\begin{tabular}[c]{@{}c@{}}Queue\\ Scene 3\end{tabular}} & \multirow{4}{*}{5}                                        & 0.75                                                     & 1                                                      & Wide                                                         & Figure~\ref{fig:queue_sim_scene3_1}  \\ \cmidrule(l){3-6} 
                                                                         &                                                           & 0.75                                                     & 1                                                      & Narrow                                                        & Figure~\ref{fig:queue_sim_scene3_2}  \\ \cmidrule(l){3-6} 
                                                                         &                                                           & 0.5                                                      & 5                                                      & Wide                                                         & Figure~\ref{fig:queue_sim_scene3_3}  \\ \cmidrule(l){3-6} 
                                                                         &                                                           & \textbf{0.5}                                                      & \textbf{5}                                                      & \textbf{Narrow}                                                        & \textbf{Figure~\ref{fig:queue_sim_scene3_4}}  \\ \bottomrule
\end{tabular}
\caption{Comparison of parameters used in the simulations of the queue behavior. Queue Scene 1 does not contain obstacles. Values in bold represent the values deemed most suitable for simulating queues.}
\label{tab:queue_params}
\end{table}

\begin{figure}[ht]
\centering
\includegraphics[width=0.45\textwidth]{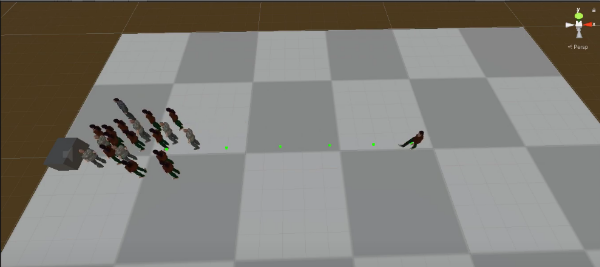}
\caption{Simulation of Queue Scene 1, considering $MarkerDensity = 0.75$, $AgentRadius = 1$, as presented in Table~\ref{tab:queue_params}.}
\label{fig:queue_sim_scene1}
\end{figure}

\begin{figure}[!htb]
  \centering
  \subfigure[fig:queueB1][Simulation of Queue Scene 2 considering $MarkerDensity = 0.75$, $AgentRadius = 1$, and a wide space between obstacles.]{\includegraphics[width=0.42\textwidth]{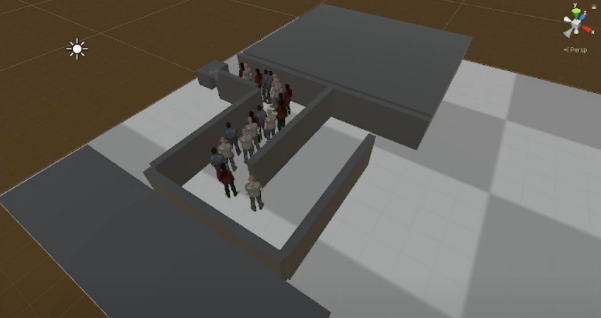}
  \label{fig:queue_sim_scene2_1}}
  \subfigure[fig:queueB2][Simulation of Queue Scene 2 considering $MarkerDensity = 0.75$, $AgentRadius = 1$, and a narrow space between obstacles.]{\includegraphics[width=0.42\textwidth]{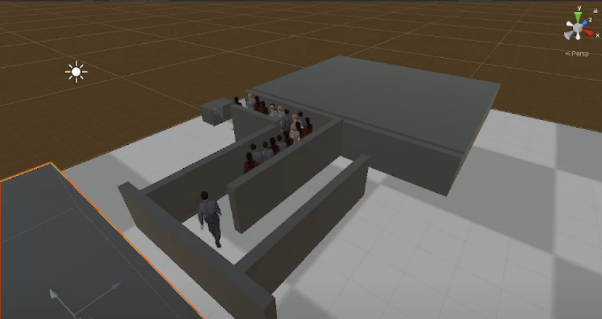}
  \label{fig:queue_sim_scene2_2}}
  \subfigure[fig:queueB3][Simulation of Queue Scene 2 considering $MarkerDensity = 0.5$, $AgentRadius = 5$, and a wide space between obstacles.]{\includegraphics[width=0.42\textwidth]{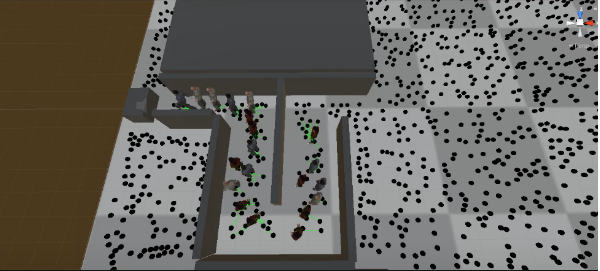}
  \label{fig:queue_sim_scene2_3}}
  \subfigure[fig:queueB4][Simulation of Queue Scene 2 considering $MarkerDensity = 0.5$, $AgentRadius = 5$, and a narrow space between obstacles.]{\includegraphics[width=0.42\textwidth]{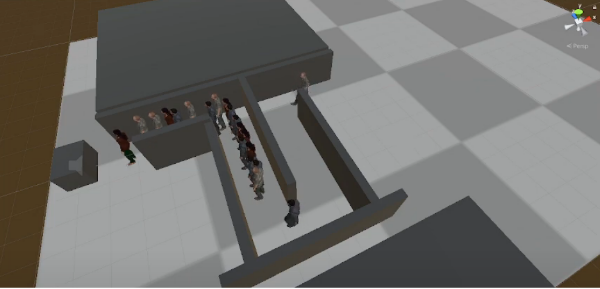}
  \label{fig:queue_sim_scene2_4}}
    \caption{Simulations of Queue Scene 2, considering the parameter configurations presented in Table~\ref{tab:queue_params}.}
    \label{fig:queue_simulation_scene2}
\end{figure}

\begin{figure}[!htb]
  \centering
  \subfigure[fig:queueC1][Simulation of Queue Scene 3 considering $MarkerDensity = 0.75$, $AgentRadius = 1$, and a wide space between obstacles.]{\includegraphics[width=0.42\textwidth]{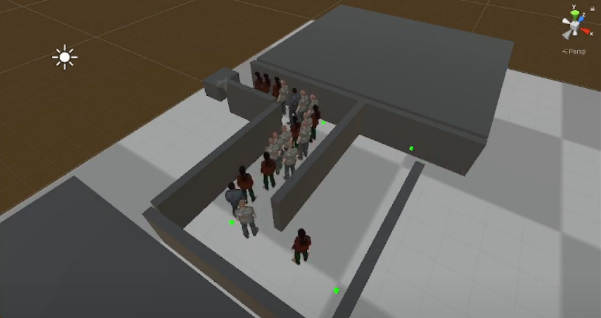}
  \label{fig:queue_sim_scene3_1}}
  \subfigure[fig:queueC2][Simulation of Queue Scene 3 considering $MarkerDensity = 0.75$, $AgentRadius = 1$, and a narrow space between obstacles.]{\includegraphics[width=0.42\textwidth]{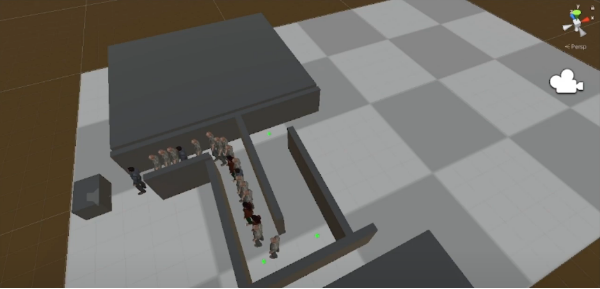}
  \label{fig:queue_sim_scene3_2}}
  \subfigure[fig:queueC3][Simulation of Queue Scene 3 considering $MarkerDensity = 0.5$, $AgentRadius = 5$, and a wide space between obstacles.]{\includegraphics[width=0.42\textwidth]{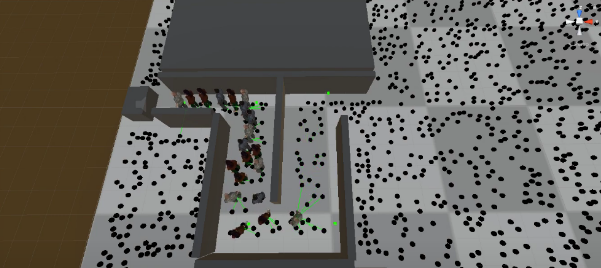}
  \label{fig:queue_sim_scene3_3}}
  \subfigure[fig:queueC4][Simulation of Queue Scene 3 considering $MarkerDensity = 0.5$, $AgentRadius = 5$, and a narrow space between obstacles.]{\includegraphics[width=0.42\textwidth]{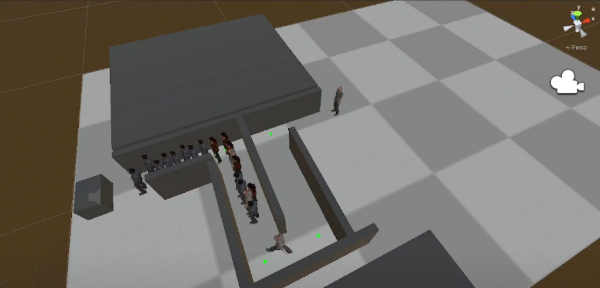}
  \label{fig:queue_sim_scene3_4}}
    \caption{Simulations of Queue Scene 3, considering the parameter configurations presented in Table~\ref{tab:queue_params}.}
    \label{fig:queue_simulation_scene3}
\end{figure}

Figure~\ref{fig:queue_sim_scene1} presents the Queue Scene 1 experiment, which solely relied on goals to guide the movement of agents. 
This experiment was conducted in a scene with a $MarkerDensity = 0.75$, $AgentRadius = 1$, and a row of 7 goals. It can be observed that this configuration did not produce an orderly queue. Some agents deviated from the desired path, moving sideways in an attempt to reach the final goal. As a result, a group of agents concentrated on the last goal, in a non-organized and non-sequential way. We hypothesize that this should be the emergent behavior of queues if people did not know how a queue should be based on social conventions.

Figure~\ref{fig:queue_simulation_scene2} presents the simulations conducted in Queue Scene 2 (Figure~\ref{fig:queue_scene_2}), where only one objective is placed at the end of the queue. Different parameter combinations were tested.
Figures~\ref{fig:queue_sim_scene2_1} and \ref{fig:queue_sim_scene2_2} present simulations with $MarkerDensity = 0.75$ and $AgentRadius = 1$, varying the available space for agents to form the queue. In Figure~\ref{fig:queue_sim_scene2_2}, the space is narrower. Figures~\ref{fig:queue_sim_scene2_3} and \ref{fig:queue_sim_scene2_4} show simulations with a $MarkerDensity = 0.5$ and $AgentRadius = 5$. This parameter combination was selected because reducing the marker density led to increased spacing between agents, requiring a larger capture radius for agents to effectively to move. In Figure~\ref{fig:queue_sim_scene2_4}, the space is reduced. In the experiments depicted in Figures~\ref{fig:queue_sim_scene2_3} and ~\ref{fig:queue_sim_scene2_4}, the queue appeared more organized, particularly when space was reduced. However, in some executions, agents became locked due to the marker positions and the lack of guiding elements within the queue.

Figure~\ref{fig:queue_simulation_scene3} presents the simulations conducted in Queue Scene 3 (Figure~\ref{fig:queue_scene_3}), where five goals are placed along the queue's path. Similar to the simulations in Queue Scene 2, we used a $MarkerDensity = 0.75$ and $AgentRadius = 1$ in Figures~\ref{fig:queue_sim_scene3_1} and \ref{fig:queue_sim_scene3_2}, also varying the available space for agents. In Figures~\ref{fig:queue_sim_scene3_3} and \ref{fig:queue_sim_scene3_4}, the values of $MarkerDensity = 0.5$ and $AgentRadius = 5$ were employed. As with Queue Scene 2, the queue appeared organized, particularly in simulations with narrower space. However, by using goals to guide agents, we managed to prevent issues with agent flow in the queue. Consequently, it was deemed that the most suitable parameter combination for simulating queues was utilized in Figure~\ref{fig:queue_sim_scene3_4}, featuring a $MarkerDensity = 0.5$, $AgentRadius = 5$, a narrow distance between obstacles, and additional goals for agent guidance.


\subsection{Implementation Details}



\section{Final Considerations}
\label{sec:final_considerations}

This work made a significant contribution by simulating structured and non-emergent behaviors in crowds, specifically focusing on rock concerts, such as \textit{moshpit} and \textit{circlepit}, which had not been adequately addressed in existing virtual human simulation literature. These behaviors were successfully implemented in the BioCrowds~\cite{BICHO201270} model through parameter modifications and the inclusion of a new repulsion behavior. Throughout the simulations, various tests were conducted, adjusting parameters to better represent the desired behaviors. The primary challenge was finding the optimal configuration for the group of agents in the BioCrowds model, allowing the behaviors to naturally emerge and facilitating interactions between agents. The results obtained were promising, demonstrating the effectiveness of the BioCrowds model in simulating structured behaviors in musical events. The simulation of these proposed behaviors uncovered new crowd movement patterns that had not been explored in the existing literature. In particular, this paper discusses how cognitive and learned behaviors can be simply modeled using an emergent crowd behavior simulation such as BioCrowds.

In future work, further analysis of the obtained results is recommended, along with comparisons between simulated behaviors and real data from actual musical shows. Additionally, exploring additional variations and complexities in crowd behaviors at musical events could prove valuable. This could involve investigating interactions between different groups of people, considering environmental factors such as lighting and acoustics, and incorporating more realistic individual behaviors, such as responses to different musical styles.


\bibliographystyle{ACM-Reference-Format}
\bibliography{bib}

\end{document}